\documentclass[apl,superscriptaddress,twocolumn,floatfix]{revtex4}
\usepackage{latexsym}
\usepackage{amsfonts}
\usepackage{amssymb}
\usepackage{amsmath}
\usepackage{graphicx}
\usepackage{natbib}

\begin{document}

\title{Phase locking of vortex based spin transfer oscillators to a microwave current}

\author{A. Dussaux}
\affiliation{Unit\'e Mixte de Physique CNRS/Thales and Universit\'e Paris Sud 11, 1 ave A. Fresnel, 91767 Palaiseau, France}
\author{A. V. Khvalkovskiy}
\affiliation{Unit\'e Mixte de Physique CNRS/Thales and Universit\'e Paris Sud 11, 1 ave A. Fresnel, 91767 Palaiseau, France}
\affiliation{A.M. Prokhorov General Physics Institute of RAS, Vavilova str. 38, 119991 Moscow, Russia} 
\author{J. Grollier}
\affiliation{Unit\'e Mixte de Physique CNRS/Thales and Universit\'e Paris Sud 11, 1 ave A. Fresnel, 91767 Palaiseau, France}
\author{V. Cros}
\affiliation{Unit\'e Mixte de Physique CNRS/Thales and Universit\'e Paris Sud 11, 1 ave A. Fresnel, 91767 Palaiseau, France}
\author{A. Fukushima}
\affiliation{National Institute of Advanced Industrial Science and Technology (AIST) 1-1-1 Umezono, Tsukuba, Ibaraki 305-8568, Japan}
\author{M. Konoto}
\affiliation{National Institute of Advanced Industrial Science and Technology (AIST) 1-1-1 Umezono, Tsukuba, Ibaraki 305-8568, Japan}
\author{H. Kubota}
\affiliation{National Institute of Advanced Industrial Science and Technology (AIST) 1-1-1 Umezono, Tsukuba, Ibaraki 305-8568, Japan}
\author{K. Yakushiji}
\affiliation{National Institute of Advanced Industrial Science and Technology (AIST) 1-1-1 Umezono, Tsukuba, Ibaraki 305-8568, Japan}
\author{S. Yuasa}
\affiliation{National Institute of Advanced Industrial Science and Technology (AIST) 1-1-1 Umezono, Tsukuba, Ibaraki 305-8568, Japan}
\author{K. Ando}
\affiliation{National Institute of Advanced Industrial Science and Technology (AIST) 1-1-1 Umezono, Tsukuba, Ibaraki 305-8568, Japan}
\author{A. Fert}
\affiliation{Unit\'e Mixte de Physique CNRS/Thales and Universit\'e Paris Sud 11, 1 ave A. Fresnel, 91767 Palaiseau, France}

\date{\today}

\begin{abstract}

Phase locking experiments on vortex based spin transfer oscillators with an external microwave current are performed. We present clear evidence of phase locking, frequency pulling, as well as fractional synchronization in this system, with a minimum peak linewidth of only 3 kHz in the locked state. We find that locking ranges of the order of 1/3 of the oscillator frequency are easily achievable because of the large tunability $\partial f/\partial I_{dc}$ observed in our vortex based systems. Such large locking ranges allow us to demonstrate the simultaneous phase locking of two independent oscillators connected in series with the external source.

\end{abstract}

\maketitle

Injection of a direct electrical current through spin-valve structures or magnetic tunnel junctions offers the possibility to induce microwave steady-state magnetization precession by the action of the spin transfer torque. Due to the magneto-resistive effects, these oscillations are converted to an a.c. electrical signal at microwave frequencies \cite{Kiselev2003,Rippard2004}. Such spin transfer nano-oscillators are advantageous for applications in wireless telecommunications, but despite significant progress in increasing the power of these oscillators and reducing the linewidth \cite{Dussaux}, these parameters however do not yet match requirements for practical applications. Synchronization of many oscillators is a solution to overcome these issues \cite{GrollierSync,Zhang}. One of the mechanisms of synchronization of many oscillators discussed in Ref. \cite{GeorgesAPL, Slavin} is based on their ability to adapt their frequency to the frequency of an injected a.c. current. Synchronization of a single oscillator to a microwave source has been demonstrated experimentally for systems in which the motion of a quasi-uniform magnetization has been excited \cite{RippardPRLinjection,GeorgesPRL}. In our work, we study synchronization to a microwave current of oscillators having a vortex in the free layer of a magnetic tunnel junction (MTJ). We have demonstrated that these \textit{spin transfer vortex oscillators} (STVOs) result in large emitted power with low linewidth compared to oscillators based on a quasi-uniform magnetization precession\cite{Dussaux}. We find that such STVOs can be locked to the fractional frequencies of the microwave current source as well as to its main frequency. The ratio of the locking range to the emission frequency can be very large, and it allows us to observe experimentally the synchronization of two separate STVOs to a common external microwave current. 

\begin{figure}[h]
   \begin{center}
    \centerline{ \includegraphics[clip,width=8 cm]{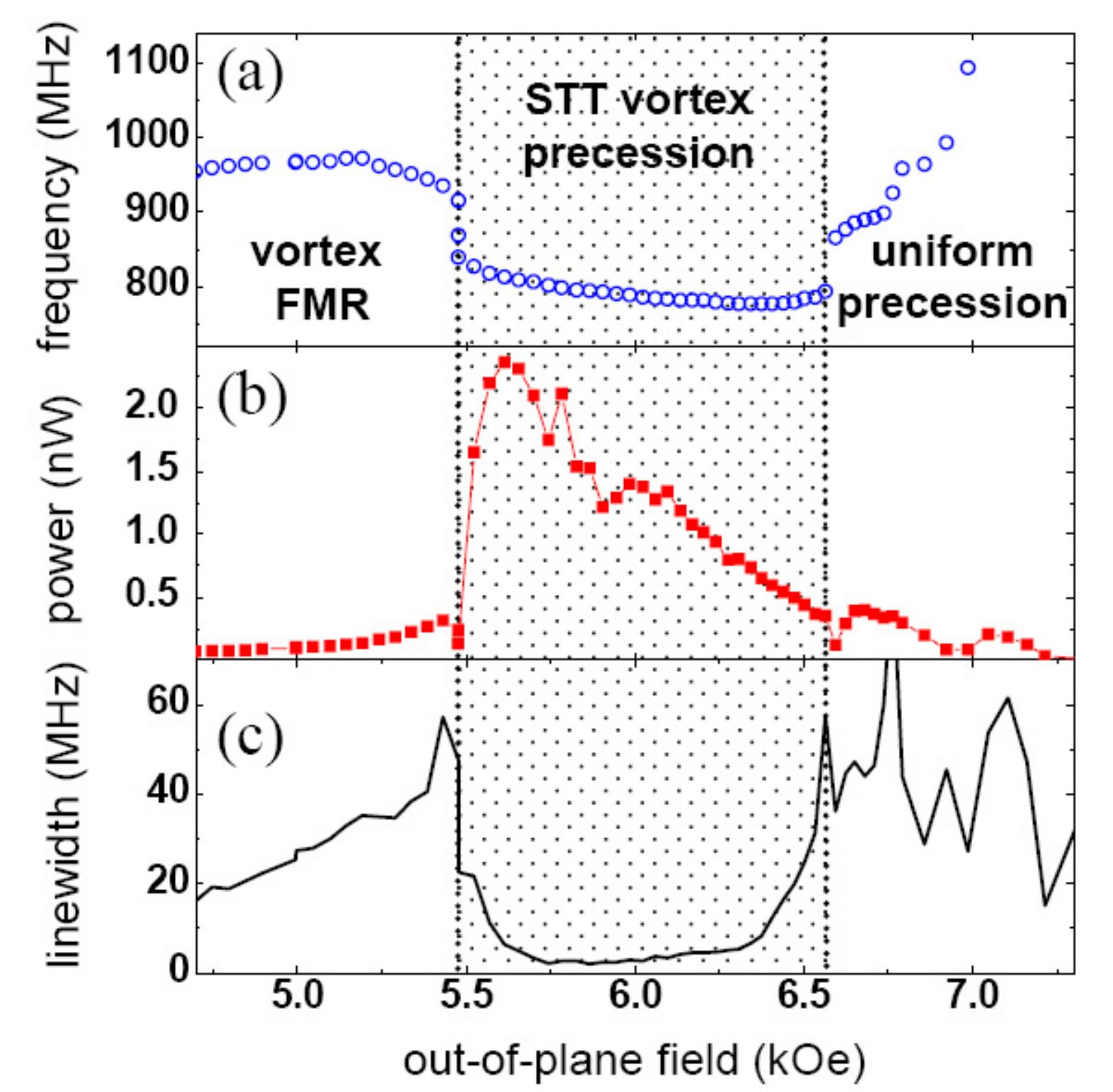}}
     \caption{(a) Frequency, (b) integrated power, (c) linewidth evolution of the emitted signal as a function of the applied out-of-plane magnetic field for $I_{dc}$ = 3.5 mA. In the dotted region, the magnetic vortex is excited by the spin transfer torque.}
    \label{fig1}
    \end{center}
\end{figure}

\begin{figure}[h]
   \begin{center}
    \centerline{ \includegraphics[clip,width=9 cm]{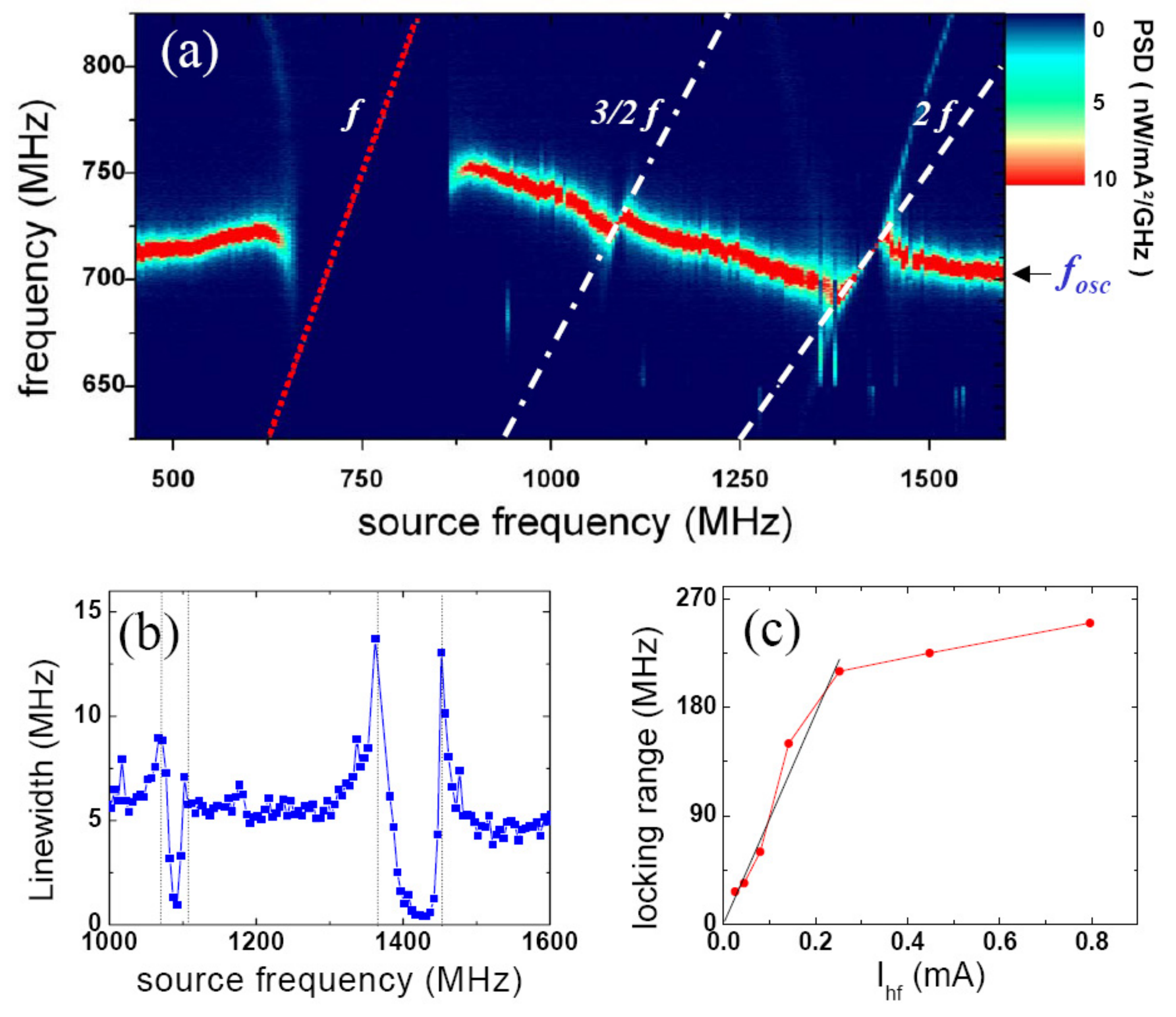}} 
        \caption{(a) Power spectrum map of the spin transfer vortex oscillator with frequency $f_{osc}$ excited by a microwave current $I_{rf}$ = 0.80 mA. The map is recorded at $H$ = + 5.76 kOe, $I_{dc}$ = 3.5 mA. The source frequency $f_{source}$ is swept from 450 MHz to 1650 MHz (red line). The white dashed lines are guides for the eyes showing 3/2 $f_{source}$ and 2 $f_{source}$. (b) linewidth of the signal emitted with an external a.c. current $I_{rf}$ = 0.80 mA swept from 1000 MHz to 1600 MHz. (c) locking range of the single vortex based oscillator as a function of the external a.c. current amplitude for $H$ =  + 5.76 kOe and $I_{dc}$ = 3.5 mA (red curve), linear fit obtained for $I_{rf}$ $\leq$ 0.25 mA (black curve). }
    \label{fig2}
    \end{center}
\end{figure}

 In this letter, we perform phase-locking experiments using STVOs made of circular shape nanopillars (diameter D = 170 nm) from the same MTJ wafer as in Ref. \cite{Dussaux}. The magnetic stacks, grown by sputtering, are composed of  PtMn 15 / CoFe 2.5 / Ru 0.85 / CoFeB 3 / MgO 1.075 / NiFe 15 / Ru 10 (nm). The ratio thickness over diameter of the free NiFe layer is chosen in such a way that a magnetic vortex is stabilized as a remanent magnetic state. The top layer of the synthetic antiferromagnet (SAF) PtMn/CoFeB/Ru/CoFeB serves as a polarizer. The tunnel magnetoresistance (TMR) is 10 $\%$. Microwave emissions are recorded on a spectrum analyzer and injection locking experiments are performed by adding a microwave circulator between the bias tee and the sample, to inject a microwave current, $I_{rf}$, from an external source. In our convention, a positive current is defined as electrons flowing from the NiFe magnetic layer to the SAF. 

As we demonstrated before \cite{Dussaux}, a rather large out-of-plane magnetic field, $H$$\approx$ 5.5 kOe, is needed in order to tilt the polarizer magnetization and therefore to induce a perpendicular component of the spin current, that is required for the excitation of the vortex gyrotropic motion with a uniform polarizer \cite{KhvalkovskiyPerp}. In Fig. \ref{fig1}, we plot the frequency (a), the power (b) and the linewidth (c) as a function of the out-of-plane field $H$ for $I_{dc}$ = 3.5 mA. For $H$ $<$ 5.5 kOe, the out-of-plane spin current is too small to excite vortex motion, thus only thermally excited vortex resonance with low power and large linewidth is observed. At very large field ($H$ $> $ 6.5 kOe), the magnetic configuration is no more a vortex but rather a quasi-uniform magnetic state with low emitted power and very large linewidth. The fields of interest for the present study go from 5.5 to 6.5 kOe, in which spin transfer torque vortex precession is present, along with large power (1.5 - 2.5 nW) and narrow linewidth (1 - 5 MHz). An important feature in this field range is that the combined action of the spin torque and the Oersted field leads to large frequency tunability $\partial f/\partial I_{dc}$ \cite{Dussaux}.

We then study the phase locking properties of our STVO for an out-of-plane field $H$ = + 5.76 kOe and $I_{dc}$ = 3.5 mA, at which the oscillator frequency (corresponding to the gyrotropic motion of the vortex core) is $ f_{osc} $ = 707 MHz, the linewidth is 4.7 MHz, and the integrated power is 2.5 nW. In Fig.\ref{fig2} (a), we present a map of the power density spectra recorded with $I_{rf}$ = 0.80 mA, as a function of the external r.f. current frequency, $f_{source}$, varying from 450 MHz to 1700 MHz (the red dotted line in Fig.\ref{fig2} (a) is the injected r.f. signal). When $f_{source}$ comes close to $f_{osc}$, the oscillator first deviates from its \textit{natural} frequency and eventually beats exactly at the source frequency. The values of $f_{source}$ for which the STVO signal disappears, define the locking range. The signal reappears when $f_{source}$ is again well separated from $f_{osc}$. These behaviors are characteristic of phase locking of a non linear oscillator to an external signal. We observe additional peaks coming from the modulation of the external rf current by the oscillator, as expected in microwave injection experiments (the peak corresponding to $f_{source}$ - $f_{osc}$ being the most visible here \cite{PuffalModul}). Note that, in the locking regime, a direct access to the signal spectral properties is impossible as the peaks from STVO and the source merge \cite{GeorgesPRL, RippardPRLinjection}. 

Interestingly, in addition to the synchronization at the fundamental frequency, we also demonstrate that the oscillator is locked to the source when its frequency $f_{source}$ is equal to some fractions of $f_{osc}$, for example $f_{source}$ $\approx$ 3/2 $f_{osc}$ and $f_{source}$ $\approx$ 2 $f_{osc}$ (dotted white lines in Fig.\ref{fig2} (a)). These effects are not related to the source non linearity since no sub-harmonics are emitted. Because of the large locking effects, the oscillator does not come back to its free running frequency between the different fractional synchronization regions. Most non-linear oscillators can exhibit such higher order synchronizations when the driving force is large enough \cite{Pikovski}. For the experiments of synchronization to a microwave current of quasi-uniform magnetization, the efficiency of the spin transfer torque did not permit to see such effects \cite{GeorgesPRL, RippardPRLinjection}. Urazhdin \textit{et al.} have recently shown that for symmetry reason, a microwave field strongly couples to the uniform mode allowing a large locking range and fractional synchronization \cite{Urazhdin}. Here, we demonstrate that with a vortex based oscillator, a microwave current turns out to be an efficient driving force.

Besides its fundamental interest, the fractional synchronization also allows us to investigate the characteristics of the emitted signal due a locked STVO. In Fig.\ref{fig2} (b), we display the variation of the signal linewidth while $f_{source}$ varies from 1.0 GHz to 1.6 GHz. Both for $f_{source}$ around $2/3$ $f_{osc}$ and $2 f_{osc}$, a strong reduction of the peak linewidth of the locked STVO is observed down a minimum value that is only due to the resolution bandwidth ($RBW =$ 470 kHz) used for this large frequency scan measurements.We have performed additional measurements at $f_{source}$ = $2f_{osc}$ with much lower RBW (0.91 kHz), allowing us to determine the intrinsic linewidth of 3 kHz. We attribute the increase of linewidth around the regions of fractional synchronization to successive locking-unlocking events occurring at the timescale of the measurements, that broaden the signal.  All the features for the linewidth visible in Fig.\ref{fig2} (b), give us definite proofs that we effectively observe phase locking of a vortex gyrotropic motion to the microwave current delivered by the external source. In particular, once locked, the small variation of the vortex oscillation frequency with time, that has been identified as the main source of the linewidth \cite{Pribiag2009}, is canceled.
  
   We have also studied the evolution of the phase locking with the amplitude of the external rf signal. In Fig.\ref{fig2} (c), we plot the locking range at the fundamental frequency i.e. $f_{source}$ = $f_{osc}$ for $I_{rf}$ ranging from 0.025 mA to 0.80 mA. For small excitations ($I_{rf}$ $\leq$ 0.25 mA), we find that the locking range increases almost linearly with $I_{rf}$ with a slope of 873 MHz/mA. In this regime of small driving force, no fractional synchronization was observed. For larger $I_{rf}$, the coupling between the oscillator and the source becomes strongly non linear thus explaining the existence of fractional synchronization regimes. A theoretical prediction of the locking range versus driving force amplitude cannot be achieved without a model for spin transfer induced vortex oscillations including non linearity in the different forces.

 \begin{figure}[h]
   \begin{center}
    \centerline{ \includegraphics[clip,width=9 cm]{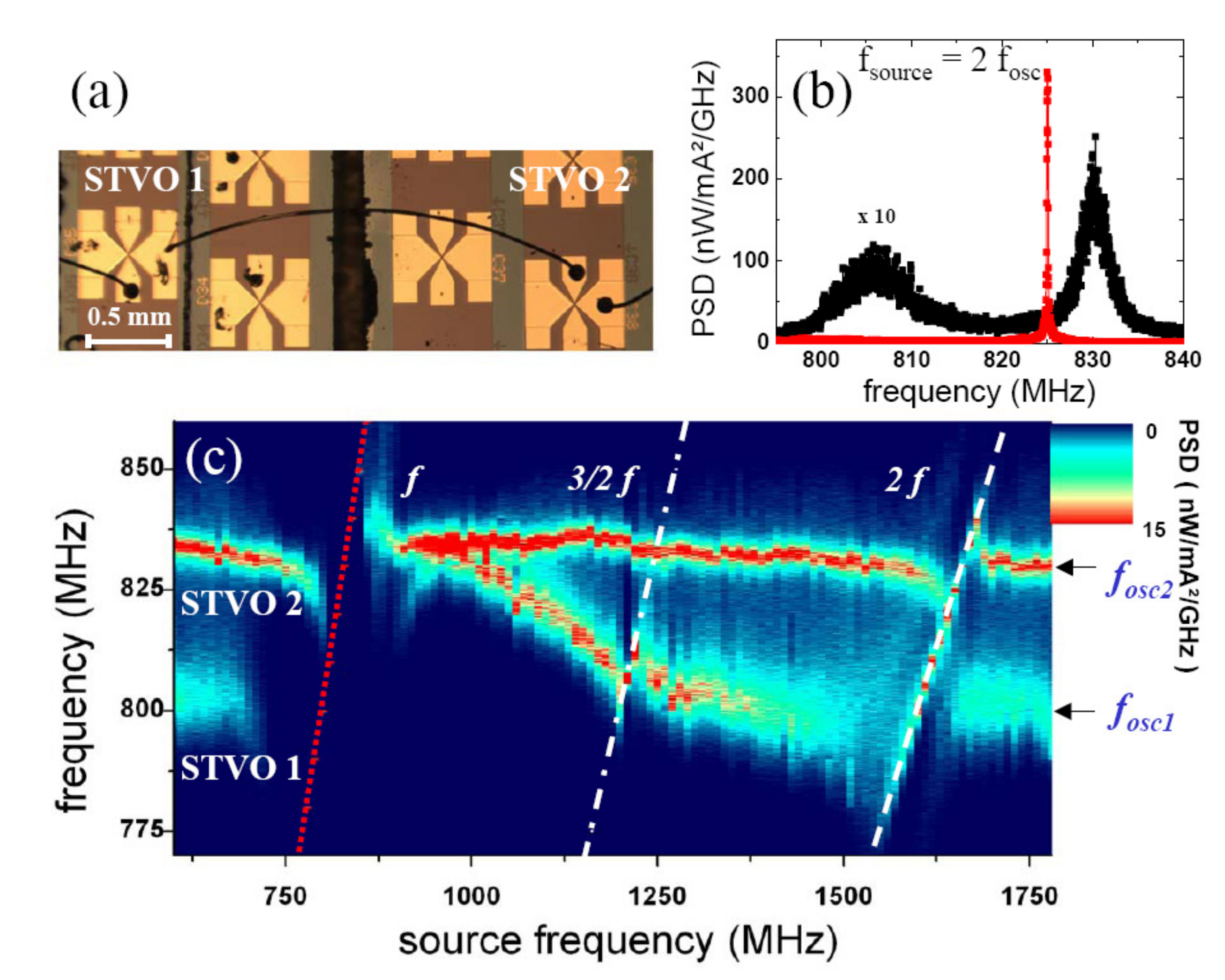}}
         \caption{(a) Optical microscope image of two separated oscillators, labelled STVOs 1 and 2, that are electrically connected in series by wire bonding. (b)  Power spectra measured for $f_{source}$ = 1650 MHz (red dots) and without source (black dots). In the latter case, the power has been multiplied by a factor of 10. (c) Power spectrum map obtained for STVO 1 and 2, with frequencies $f_{osc1}$ and $f_{osc2}$, connected in series  recorded with $H$ = + 5.82 kOe, $I_{dc}$ = 3.5 mA and sweeping the source frequency $f_{source}$ from 600 MHz to 1800 MHz with a microwave current $I_{rf}$ = 0.67 mA. The white dashed lines show the evolution of 3/2 $f_{source}$ and 2 $f_{source}$.}
    \label{fig3}
    \end{center}
\end{figure}

In previous studies, the maximum locking range obtained by microwave current injection in the quasi-uniform magnetization regime, represented only a very small fraction (about $1\%$ ) of their free running frequency \cite{GeorgesPRL, RippardPRLinjection}. In contrast, we demonstrate here that the locking range with our STVO goes up to 250 MHz for $I_{rf}$ = 0.80 mA, that is more than 35\% of the oscillator gyrotropic frequency. 
Key parameters which define the oscillator ability to synchronize to an external signal are the linewidth and the tunability, i.e. $\partial f/\partial I_{dc}$ \cite{GeorgesPRL,RippardPRLinjection}. For this reason, the experiment presented in Fig.\ref{fig2} has been performed at the field (5.76 kOe) which gives the highest tunability to our STVO: 160 MHz/mA. This large value is in striking contrast with the small tunabilities reported on STVOs by other groups, typically a few 10 MHz/mA \cite{Pribiag, Pufall, Mistral}.
In our system the spin transfer emissions occur in a field range where the resonant frequency is strongly affected by the external field (see supplementary information in Ref. \cite{Dussaux}). The out-of-plane field deforms the vortex shape whereas torques due to spin transfer and Oersted fields tend to push the magnetization back in-plane, leading to strong frequency variations with current. In contrast, for small fields ($H$ $<$ 5.5 kOe), the tunability is less than 40 MHz/mA resulting in a small locking range of about 40 MHz. In the field range of large power spin transfer vortex precession of Fig. \ref{fig1}, the large locking-range of 250 MHz results from the combination of high tunability and small signal linewidth. 

The large locking range of our STVOs and the small frequency dispersion from sample to sample provide us with the opportunity to demonstrate coherent oscillations of two independent oscillators locked to the source. Subsequently, we connect by wire bonding two oscillators in series, labeled STVO1 and STVO2, separated by a few millimeters (see Fig.\ref{fig3} (a)). As expected, the emission spectrum contains two independent peaks, corresponding to the emission of each oscillator. The spectral maxima are at $f_{osc1}$ = 811 MHz and $f_{osc2}$ = 832 MHz, the linewidths of the peaks are 7.0 MHz and 3.3 MHz, and the integrated power are 0.6 nW and 1.0 nW respectively (see black curve in Fig.\ref{fig3} (b)). We then study the locking of these two STVOs to the source, for an external field $H$ =  + 5.82 kOe, $I_{dc}$ = 3.5 mA by injecting, as before, a microwave current $I_{rf}$ of 0.67 mA. In Fig. \ref{fig3} (c), we show the map of the power spectrum for this system with two STVOs. We clearly observe for the two oscillators both synchronization at the fundamental frequency as well as fractional synchronization to the external source. When the two  locking ranges overlap, as for example, for 800 MHz $<f_{source}<$ 860 MHz, both oscillators are phase locked to the source, and emit in-phase. The regions of synchronization to the second harmonic do also overlap, allowing us to clearly demonstrate that the emission occurs at a single frequency with a strong narrowing of the peak (see red curve in Fig.\ref{fig3}b at $f_{source}$ = 1650 MHz). 

In summary, we have demonstrated that a spin transfer vortex oscillator can be locked to an external rf current not only at its main frequency but also at fractional frequencies such as $3/2 f_{osc}$ or $2 f_{osc}$. Inducing large tunabilities, $\partial f/\partial I_{dc}$, of the vortex gyrotropic mode by using appropriate dc current and out of plane field values, we achieve large locking ranges, of the order of the STVO frequency. The observation of higher order synchronization allows to study directly the power spectrum characteristics of the locked STVOs, for which we find for example a linewidth as low as 3 kHz. In addition, we have shown experimentally the coherent oscillations of two STVOs connected in series synchronized to the external source. Our results demonstrate that vortex based spin transfer nano-oscillators are good candidates to achieve the synchronization of a large array of oscillators through their self-emitted microwave currents.

The authors acknowledge Y. Nagamine, H. Maehara and K. Tsunekawa of CANON ANELVA for preparing the MTJ films. Financial support by the CNRS and the ANR agency (VOICE PNANO-09-P231-36) and EU grant (MASTER No. NMP-FP7-212257) is acknowledged. A.V.K. is partially supported by the RFBR (Grant No. 09-02-01423).

\end{document}